\documentclass[%
 reprint,
 amsmath,amssymb,
 aps,
]{revtex4-2}
\usepackage{mathptmx}
\usepackage{etoolbox}
\usepackage{xcolor}
\usepackage{float}
\usepackage{hyperref}
\usepackage{graphicx}
\usepackage{dcolumn}
\usepackage{bm}

\begin{document}

\preprint{APS/123-QED}

\title{Stacking-Dependent Magnetism and Tunable Half-Metallicity in Bilayer Janus 1T-MnSSe}

\author{Jakkapat Seeyangnok$^{1}$}
 \email{jakkapatjtp@gmail.com} 
\author{Udomsilp Pinsook$^{1}$}%
 \email{Udomsilp.P@Chula.ac.th}

\author{Graeme J Ackland$^{2}$}
 \email{gjackland@ed.ac.uk} 
\affiliation{$^{1}$Department of Physics, Faculty of Science, Chulalongkorn University, Bangkok, Thailand.\\
$^{2}$Centre for Science at Extreme Conditions, School of Physics and Astronomy, University of Edinburgh, Edinburgh, United Kingdom}%


\date{\today}

\begin{abstract}
We investigate the structural, electronic, and magnetic properties of bilayer Janus 1T-MnSSe using first-principles calculations. Various AA- and AB-type stacking configurations are considered to examine the influence of interlayer registry on magnetic ordering and exchange interactions. The nonmagnetic state is unstable for all stackings, confirming intrinsic magnetism. The AA2 stacking is identified as the ground state and exhibits A-type antiferromagnetic ordering, indicating antiferromagnetic interlayer coupling. Monte Carlo simulations based on an effective Ising model reveal enhanced magnetic transition temperatures in the bilayer relative to the monolayer, with N\'eel temperatures above 300~K for antiferromagnetic stackings and Curie temperatures up to 250~K for ferromagnetic phases. Several stacking configurations exhibit robust half-metallic ferromagnetism with nearly 100\% spin polarization at the Fermi level. Moreover, the half-metallic state can be tuned and ultimately transformed into a metallic ferromagnetic phase through carrier doping and biaxial strain. These findings establish bilayer MnSSe as a promising platform for controllable interlayer magnetism and spintronic applications in two-dimensional materials.
\end{abstract}

\keywords{bilayer Janus TMDs; 1T-MnSSe; two-dimensional magnetism; magnetic stacking; interlayer exchange coupling; antiferromagnetism; density functional theory}
\maketitle

\section{Introduction}
Two-dimensional (2D) magnetic materials have attracted significant attention due to their fundamental interest~\cite{fert2008nobel,wolf2001spintronics} and potential applications in next-generation technologies~\cite{bhimanapati2015recent,novoselov20162d,mas20112d,schaibley2016valleytronics}, particularly in spintronic devices~\cite{dery2007spin,li2016first}. The experimental realization of intrinsic magnetism in atomically thin systems, such as CrI$_3$ and Cr$_2$Ge$_2$Te$_6$~\cite{huang2017layer,gong2017discovery}, has opened new opportunities to explore magnetic ordering, interlayer exchange interactions, and external tunability via stacking, strain, and electric fields. In this context, van der Waals layered materials provide a versatile platform in which magnetic properties can be engineered through layer number and stacking geometry~\cite{qiao2024tunable,burch2018magnetism,sivadas2018stacking}.

Transition-metal dichalcogenides (TMDs) represent an important class of 2D materials exhibiting diverse electronic ground states, including charge density wave (CDW) phases~\cite{zhu2017misconceptions,johannes2008fermi,silva2016electronic,dalal2025flat} and magnetism~\cite{qiao2024tunable,bonilla2018strong,wang2020bethe,li2020coupling}. Among them, Janus TMDs—formed by breaking out-of-plane mirror symmetry via asymmetric chalcogen functionalization—have attracted increasing interest. This structural asymmetry gives rise to novel physical phenomena, including enhanced superconductivity~\cite{seeyangnok2026tunable,seeyangnok2026theoretical,seeyangnok2024superconductivity_wseh,seeyangnok2024superconductivity_wsh} and tunable magnetic interactions~\cite{he2018two,chen2022electrically,kulish2017single}. In addition, the interplay between competing electronic phases, driven by a high density of states, has been reported in these systems~\cite{seeyangnok2026competition,seeyangnok2026moxhcdw,seeyangnok2025competition}. When magnetic transition-metal elements are incorporated, Janus TMDs offer a promising route to realize controllable 2D magnetism~\cite{seeyangnok2026competition,sukserm2025half,qiu2024high}.

A broad range of 2D magnetic materials has been explored, including transition-metal halides, MXenes, metal phosphorus trichalcogenides, and TMD-based systems~\cite{burch2018magnetism,gibertini2019magnetic,gong2019two}. These materials exhibit diverse magnetic behaviors, such as ferromagnetism, antiferromagnetism, and half-metallicity~\cite{bonilla2018strong}, governed by reduced dimensionality, strong electron correlations, and spin–orbit coupling~\cite{gibertini2019magnetic}. In particular, manganese-based systems have attracted considerable attention due to their robust magnetic moments originating from partially filled $d$ orbitals~\cite{he2018two,chen2022electrically}.

In this context, the magnetic properties of Janus TMDs have been extensively investigated~\cite{he2018two}, revealing intriguing features such as half-metallicity in MnSSe. Motivated by its potential applications, the MnSSe monolayer has been further studied~\cite{chen2022electrically}. These works demonstrate that Mn-based TMDs can host stable magnetic moments and tunable electronic structures. In bilayer systems, however, the magnetic ground state is strongly influenced by interlayer coupling, which depends sensitively on stacking configuration and interlayer distance~\cite{sivadas2018stacking,li2020coupling,anderson1950antiferromagnetism,goodenough1955theory}. Variations in stacking can significantly modify orbital overlap and exchange pathways, leading to competing ferromagnetic and antiferromagnetic states. Despite these advances, a systematic understanding of stacking-dependent magnetic phases in bilayer Janus TMDs remains lacking.

In this work, we present a comprehensive first-principles study of bilayer 1T-MnSSe, focusing on the interplay between stacking geometry and magnetic ordering. Multiple AA- and AB-type stacking configurations are considered, along with several magnetic states, including ferromagnetic (FM), A-type antiferromagnetic (AAF), and G-type antiferromagnetic (GAF) orders. By comparing total energies, we identify the magnetic ground state and elucidate the role of interlayer exchange interactions. Our results reveal that the AA2 stacking with AAF order is energetically favored, highlighting the critical role of stacking engineering in controlling magnetism in bilayer Janus materials. These findings provide valuable insights for the design of 2D magnetic heterostructures based on Janus TMDs.
\section{Computational Methods}
All calculations were performed using density functional theory (DFT) as implemented in the Quantum Espresso package~\cite{giannozzi2009quantum}. Projector augmented-wave (PAW) pseudopotentials were employed together with the Perdew--Burke--Ernzerhof (PBE) form of the generalized gradient approximation (GGA)~\cite{perdew1996generalized}. A plane-wave energy cutoff of 80~Ry for the wave functions and 320~Ry for the charge density was used. The Brillouin-zone sampling was carried out using Monkhorst--Pack $k$-point meshes~\cite{monkhorst1976special} of $23 \times 23 \times 1$ and $11 \times 11 \times 1$ for the primitive cell and the $2 \times 2 \times 1$ supercell, respectively. A vacuum spacing of 20~\AA\ was applied along the out-of-plane direction to avoid spurious interactions between periodic images. The Methfessel--Paxton smearing method~\cite{methfessel1989high} with a degauss value of 0.02~Ry was employed to describe the Fermi-level occupation. To include on-site Coulomb interactions, the DFT+$U$ approach was applied with an effective Hubbard parameter of $U = 3.0$~eV. In the bilayer case, van der Waals corrections~\cite{grimme2010consistent} were included to account for interlayer interactions.

To investigate the thermal stability of the ferromagnetic order, we performed classical Monte Carlo simulations based on an effective Ising Hamiltonian,
\begin{equation}\label{MC_ML}
\mathcal{H} = - J \sum_{\langle i,j \rangle} S_i S_j ,
\end{equation}
where $S_i=\pm1$ denotes the Ising spin variable at site $i$, and the summation runs over nearest-neighbor pairs without double counting where the exchange coupling constant $J$ was extracted from first-principles total-energy calculations by mapping the DFT total energies of different magnetic configurations onto the Ising model.

The exchange coupling constant $J$ was extracted from first-principles total-energy calculations by mapping the DFT total energies of different magnetic configurations onto the Ising model. For the triangular lattice of Mn atoms, each Mn has six nearest neighbors, corresponding to three independent exchange bonds per formula unit. By explicitly counting the bond contributions in the ferromagnetic (FM) and antiferromagnetic configurations, the total energies per formula unit can be written as
\begin{equation}
E_{\mathrm{FM}} = -3 J M^2, \qquad 
E_{\mathrm{AFM}} = + J M^2 ,
\end{equation}
which leads to
\begin{equation}
\Delta E = E_{\mathrm{AFM}} - E_{\mathrm{FM}} = 4 J M^2 .
\end{equation}
Accordingly, the exchange coupling constant is obtained as
\begin{equation}
J = \frac{E_{\mathrm{AFM}} - E_{\mathrm{FM}}}{4 M^2},
\end{equation}
where $M=3.0~\mu_B$ is the magnetic moment per Mn atom and $E_{\mathrm{AFM}} - E_{\mathrm{FM}} = 82~\mathrm{meV}$/unit cell.

The finite-temperature magnetic properties of the MnSSe bilayer are described using an effective bilayer Ising Hamiltonian that captures both intralayer and interlayer exchange interactions,
\begin{equation}
\mathcal{H}
=
- J \sum_{\langle i,j \rangle,\, l} S_{i,l} S_{j,l}
- J' \sum_{i} S_{i,1} S_{i,2},
\end{equation}
where $S_{i,l}=\pm1$ denotes the Ising spin at site $i$ in layer $l=1,2$. The first term represents the nearest-neighbor intralayer exchange interaction $J$, which favors ferromagnetic alignment within each monolayer, while the second term describes the interlayer exchange interaction $J'$ between corresponding sites in the two layers.

The exchange parameters are extracted from first-principles total-energy calculations by mapping the DFT energies of different magnetic configurations onto the Ising model. The energy difference is defined relative to the magnetic ground state (GS),
\begin{equation}
\Delta E_X = E_X - E_{\mathrm{GS}} .
\end{equation}

Since the DFT total energies are expressed per formula unit, each formula unit contains one interlayer Mn--Mn pair. The energy difference between two configurations that differ only in their interlayer alignment therefore satisfies
\begin{equation}
\Delta E = 2 J' M^2,
\end{equation}
where $M$ is the magnetic moment per Mn atom used in the Ising mapping.

If the ferromagnetic (FM) configuration is the ground state, the interlayer exchange parameter is obtained from the energy difference between the A-type antiferromagnetic (AAF) and FM states,
\begin{equation}
J' = \frac{E_{\mathrm{AAF}} - E_{\mathrm{FM}}}{2 M^2},
\end{equation}
which yields $J' > 0$, favoring parallel interlayer alignment.

If instead the A-type antiferromagnetic (AAF) configuration is the ground state, all energy differences are referenced to $E_{\mathrm{AAF}}$, such that
\begin{equation}
\Delta E_{\mathrm{FM}} = E_{\mathrm{FM}} - E_{\mathrm{AAF}},
\end{equation}
and the interlayer exchange parameter becomes
\begin{equation}
J' = \frac{E_{\mathrm{FM}} - E_{\mathrm{AAF}}}{2 M^2},
\end{equation}
which yields $J' < 0$, corresponding to antiparallel interlayer coupling.

The G-type antiferromagnetic (GAF) configuration involves additional intralayer spin reversals and is therefore primarily used to extract the intralayer exchange parameter $J$, while the FM--AAF energy difference isolates the interlayer exchange interaction $J'$.

This effective bilayer Ising model forms the basis for subsequent Monte Carlo simulations, which are employed to determine the Curie temperature ($T_C$) for ferromagnetic stackings and the Néel temperature ($T_N$) for antiferromagnetic stackings.
\section{Results and Discussion}

\subsection{Structures and Magnetic configurations}
    \begin{figure*}[ht]
        \centering
        \includegraphics[width=17cm]{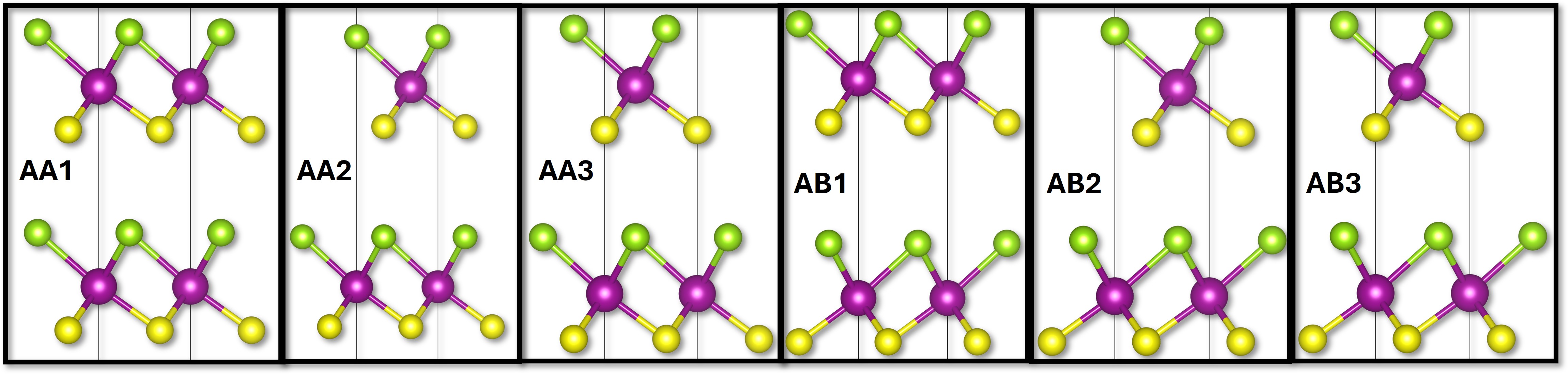}
        \caption{Top and side views of the considered primitive cell of bilayer stacking configurations. The AA-type stackings are labeled as AA1, AA2, and AA3, while the AB-type stackings are labeled as AB1, AB2, and AB3. Different relative atomic registries between the two layers are shown to illustrate the possible stacking arrangements considered in this work.}
        \label{fig:bilayer_stackings}
    \end{figure*}

    The hexagonal 1T-MnSSe crystallizes in the trigonal space group $P\overline{3}m1$ (No.~156). In this structure, the transition-metal atom Mn occupies the Wyckoff position $(0,0)$, while the chalcogen atoms Se and S are located at the Wyckoff positions $(2/3,\,1/3)$ and $(1/3,\,2/3)$, respectively, in the $x$--$y$ plane, defining the in-plane lattice constant $a$. The possible bilayer stacking configurations are shown in Fig.~\ref{fig:bilayer_stackings}, including three AA-type stackings (AA1, AA2, and AA3) and three AB-type stackings (AB1, AB2, and AB3), which arise from different relative lateral shifts between the two layers.

    \begin{figure}[h!]
        \centering
        \includegraphics[width=8.5cm]{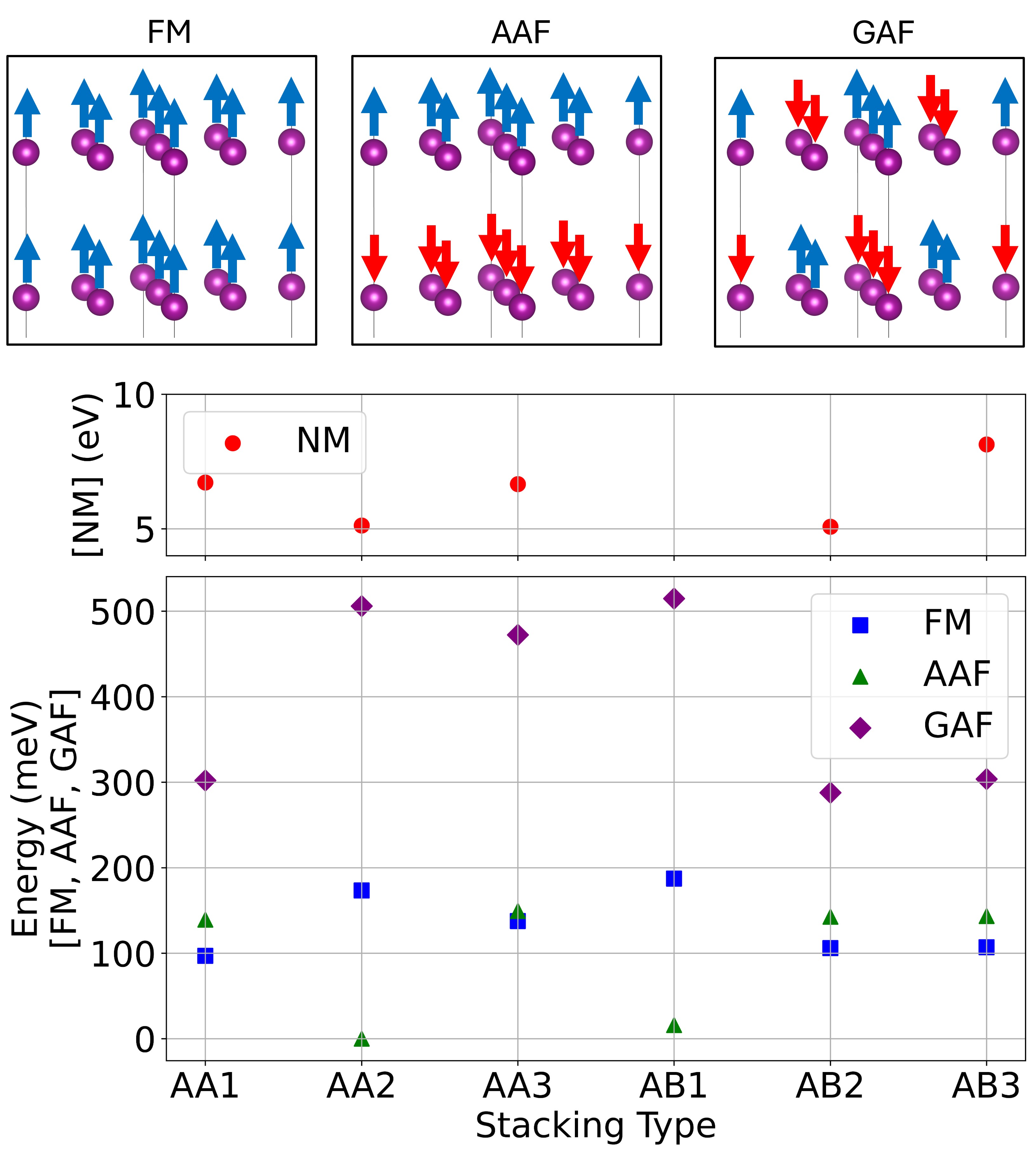}
        \caption{(a) Schematic 2$\times$2$\times$1 supercell illustration of the considered magnetic configurations: ferromagnetic (FM), A-type antiferromagnetic (AAF), and G-type antiferromagnetic (GAF) orders. Blue and red arrows indicate spin-up and spin-down orientations, respectively, on the magnetic atoms. (b) Relative total energies of the nonmagnetic (NM), FM, AAF, and GAF states for different bilayer stacking configurations. The energies are given relative to the lowest-energy magnetic state for each stacking.}
        \label{fig:magnetic_configs}
    \end{figure}

    Fig.~\ref{fig:magnetic_configs} illustrates the magnetic configurations considered in this work. We examined the ferromagnetic (FM) state, in which all Mn magnetic moments are aligned parallel, as well as two antiferromagnetic arrangements, namely A-type antiferromagnetic (AAF) and G-type antiferromagnetic (GAF) orders. In the AAF configuration, the magnetic moments are aligned ferromagnetically within each layer but are antiferromagnetically coupled between the layers. In contrast, the GAF configuration corresponds to antiferromagnetic coupling both within and between the layers. These magnetic configurations were used to evaluate the relative magnetic stability and to extract the interlayer exchange interactions.

    The relative total energies of the different magnetic phases for each bilayer stacking configuration are summarized in Fig.~\ref{fig:magnetic_configs}. For all stackings, the nonmagnetic (NM) state is significantly higher in energy, confirming the magnetic nature of bilayer 1T-MnSSe. Among the magnetic configurations, clear energy differences are observed between the FM, AAF, and GAF states, indicating a strong dependence of magnetic stability on both stacking geometry and spin arrangement. Notably, the AAF configuration in the AA2 stacking exhibits the lowest total energy among all considered cases, identifying AA2--AAF as the magnetic ground state of the bilayer system. This result suggests that antiferromagnetic interlayer coupling is energetically favored for this stacking, which plays an important role in determining the interlayer exchange interaction.

\begin{table*}[ht]
\caption{Calculated structural, electronic, and magnetic properties of monolayer (ML) and bilayer (BL) MnSSe with different stacking configurations. The lattice constant $a$ and TM--TM bond length $d_{\mathrm{TM-TM}}$ are given in \AA. G.S. denotes the ground states of electronic configuration. $J$ and $J'$ represent the intralayer and interlayer magnetic exchange interactions, respectively. $T_c$ is the Curie ( or N\'eel) temperature (K), and $M_{\mathrm{tot}}$ is the total magnetic moment ($\mu_B$).}
\label{tab:structure}
\begin{ruledtabular}
\begin{tabular}{lccccccc}
Structure & $a$ (\AA) & $d_{\mathrm{TM-TM}}$ (\AA) & G.S. & $J$ (meV) & $J'$ (meV) & $T_c$ & $M_{\mathrm{tot}}$ ($\mu_B$) \\
\hline
ML        & 3.55 & -    & HM-FM & 2.28 & -     & 190 & 3.00 \\
BL-AA1    & 3.59 & 5.41 & HM-FM & 2.28 & 2.36  & 250 & 6.00 \\
BL-AA2    & 3.73 & 5.19 & AAF   & 2.28 & -9.63 & 330 & 0.04 \\
BL-AA3    & 3.59 & 5.73 & HM-FM & 2.28 & 0.66 & 220 & 5.99 \\
BL-AB1    & 3.73 & 5.20 & AAF   & 2.28 & -9.51 & 320 & 0.07 \\
BL-AB2    & 3.60 & 5.52 & HM-FM & 2.28 & 2.05 & 250 & 5.99 \\
BL-AB3    & 3.59 & 5.50 & HM-FM & 2.28 & 2.03    & 250 & 5.99 \\
\end{tabular}
\end{ruledtabular}
\end{table*}

\subsection{Half-Ferromagnetic MnSSe Monolayer}

\begin{figure*}[ht]
    \centering
    \includegraphics[width=17cm]{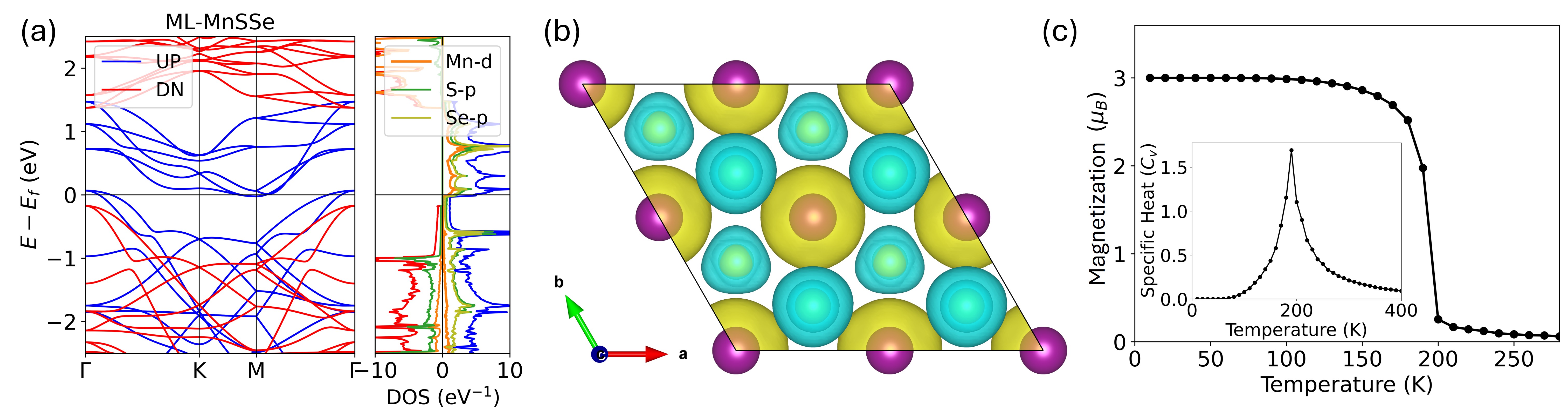}
    \caption{
    (a) Spin-resolved electronic band structure and projected density of states (PDOS) of the ML-MnSSe monolayer, where blue and red curves denote the spin-up and spin-down channels, respectively. 
    (b) Real-space spin polarization, $\rho_{\uparrow}-\rho_{\downarrow}$. Purple spheres represent Mn atoms. The yellow (cyan) isosurfaces denote positive (negative) spin polarization, corresponding to spin-up (spin-down) spin density.
    (c) Temperature dependence of the magnetic moment $M(T)$ and the specific heat $C_V(T)$ obtained from Monte Carlo simulations on a $100\times100$ lattice with $10^8$ Monte Carlo steps per temperature. The Curie temperature is estimated to be $T_C \approx 190$ K from the peak in $C_V$.
    }
    \label{fig:mn_sse}
\end{figure*}

    Fig.~\ref{fig:mn_sse} summarizes the electronic structure and finite-temperature magnetic properties of the ML-MnSSe monolayer. To establish the magnetic ground state, we compared the total energies of several magnetic configurations and found that the ferromagnetic (FM) state is energetically most favorable. The nonmagnetic (NM) configuration lies approximately $2.4$~eV higher in energy than the FM state, while the collinear antiferromagnetic (CAF) and G-type antiferromagnetic (GAF) states are higher by $88$~meV and $82$~meV, respectively. These substantial energy differences unambiguously confirm the FM ordering as the ground state of the ML-MnSSe monolayer. Consistent with this result, the spin-resolved band structure and projected density of states shown in Fig.~\ref{fig:mn_sse}(a) exhibit a clear exchange splitting between the spin-up and spin-down channels. The electronic states near the Fermi level are predominantly derived from Mn $d$ orbitals with additional contributions from chalcogen $p$ states, indicating that the magnetic behavior is mainly governed by Mn-derived electronic states.

    The real-space spin polarization distribution, defined as $\rho_{\uparrow}-\rho_{\downarrow}$, is displayed in Fig.~\ref{fig:mn_sse}(b). The spin density is strongly localized around the Mn atoms, while only weak and oppositely polarized spin density is induced on the surrounding S and Se atoms. The total magnetic moment of the system is $3.0~\mu_B$ per unit cell, which is primarily contributed by the Mn atom with a local magnetic moment of $3.5~\mu_B$. In contrast, the Se and S atoms carry smaller antiparallel magnetic moments of $-0.3~\mu_B$ and $-0.2~\mu_B$, respectively. This magnetic moment distribution confirms that the magnetism is dominated by the Mn sublattice, with antiferromagnetically polarized chalcogen atoms arising from hybridization effects, consistent with the orbital-resolved electronic structure.

    The Monte Carlo simulations based on based on an effective Ising Hamiltonian of Eqn.~\ref{MC_ML} were carried out on a $100\times100$ lattice using $10^{8}$ Monte Carlo steps per temperature to ensure statistical convergence. Fig.~\ref{fig:mn_sse}(c) presents the temperature dependence of the magnetization $M(T)$ and the specific heat $C_V(T)$. As the temperature increases, the magnetization decreases continuously and vanishes at the magnetic phase transition, while the specific heat exhibits a pronounced peak associated with enhanced energy fluctuations.

    The Curie temperature is estimated to be $T_C \approx 190$~K from the maximum of the specific heat which is consistent with previous studies~\cite{he2018two,chen2022electrically}. Although finite-size effects lead to a rounded transition, the pronounced peak in $C_V$ together with the simultaneous collapse of the magnetization clearly indicates a ferromagnetic–paramagnetic phase transition. These results demonstrate that monolayer MnSSe hosts robust long-range ferromagnetic order above liquid-nitrogen temperature, highlighting its potential for two-dimensional spintronic applications.    
\subsection{Magnetic Configurations of the van der Waals MnSSe Bilayer}
\begin{figure*}[ht]
    \centering
    \includegraphics[width=17cm]{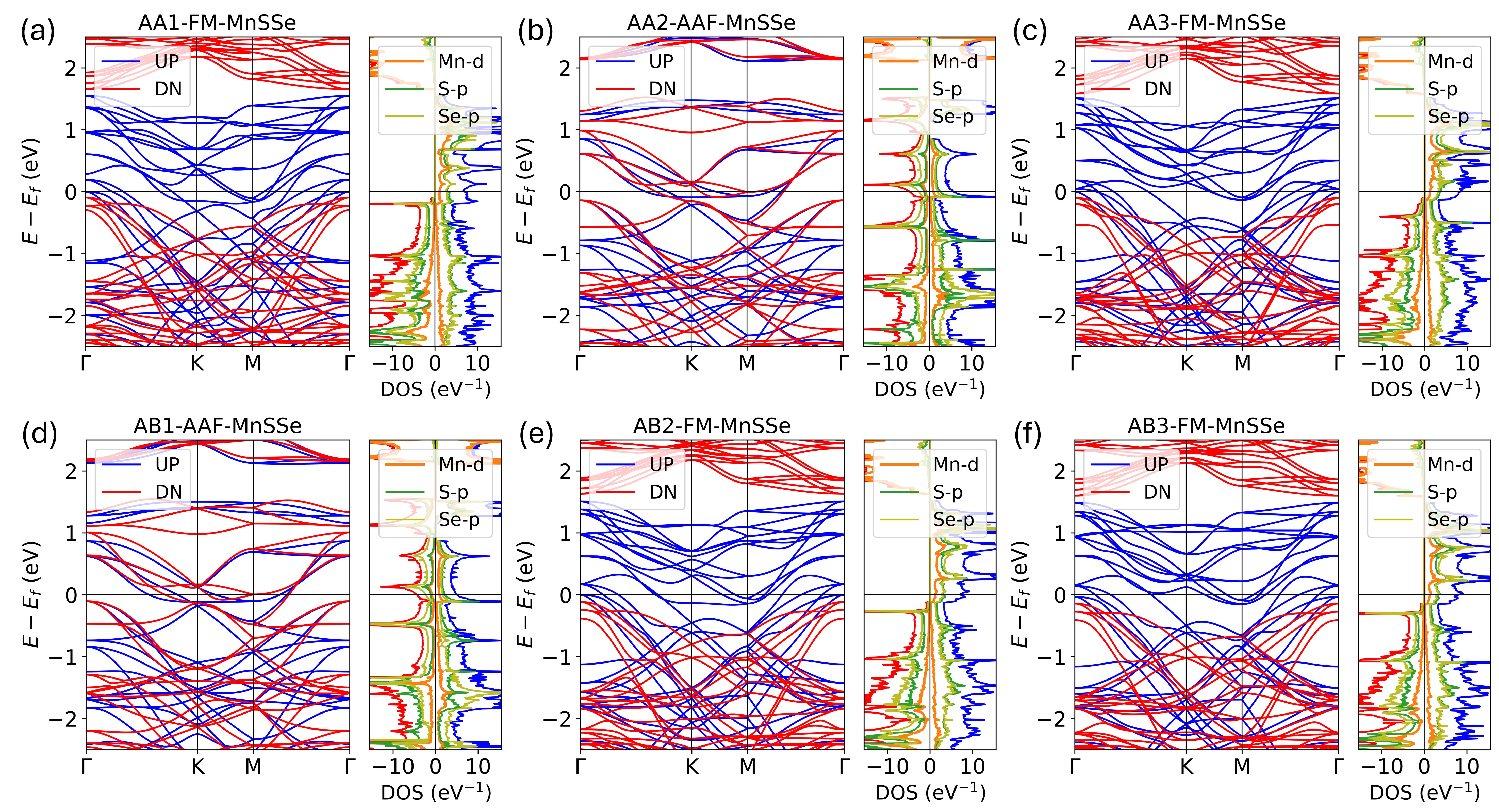}
    \caption{
    Spin-resolved electronic band structures and corresponding projected density of states (PDOS) of van der Waals bilayer Janus MnSSe for different stacking configurations.
    Panels (a), (c), (e), and (f) correspond to the half-metallic ferromagnetic (FM) states of AA1, AA3, AB2, and AB3 stackings, respectively, where one spin channel is metallic while the opposite spin channel exhibits a band gap at the Fermi level.
    Panels (b) and (d) show the AA2 and AB1 stackings, which stabilize an A-type antiferromagnetic (AAF) ground state characterized by ferromagnetic ordering within each monolayer and antiferromagnetic coupling between the two layers.
    Blue and red curves denote the spin-up and spin-down channels, respectively.
    }
    \label{fig:band_dos_mnsse}
    \end{figure*}
    In bilayer magnetic systems, the relative alignment of magnetic moments between layers introduces additional degrees of freedom beyond those present in monolayers, leading to a rich variety of possible magnetic ground states. In particular, the competition between intralayer and interlayer exchange interactions can stabilize distinct spin configurations depending on the stacking geometry and interlayer coupling strength. For the MnSSe bilayer, several magnetic polarization patterns are possible, including ferromagnetic (FM), A-type antiferromagnetic (AAF), and G-type antiferromagnetic (GAF) orderings, which differ in the relative spin alignment within and between layers.

    In this subsection, we systematically examine these competing magnetic configurations for different bilayer stacking arrangements. By comparing their total energies obtained from first-principles calculations, we identify the magnetic ground state for each stacking and quantify the energetic stability of the corresponding polarization. This analysis provides the foundation for understanding the interlayer exchange interactions in the MnSSe bilayer and motivates the subsequent investigation of finite-temperature magnetic properties.

The spin-resolved electronic band structures and projected density of states (PDOS) shown in Fig.~\ref{fig:band_dos_mnsse} reveal a pronounced stacking dependence of both the magnetic ordering and the electronic character of the MnSSe bilayer. For the AA1, AA3, AB2, and AB3 stacking configurations as shown in Fig.~\ref{fig:band_dos_mnsse}(a), (c), (e), and (f), the system stabilizes a ferromagnetic ground state with a half-metallic electronic structure, in which one spin channel remains metallic while the opposite spin channel exhibits a finite band gap at the Fermi level. This behavior implies nearly 100\% spin polarization at the Fermi energy, which is a desirable characteristic for spin-filtering and spin-injection applications.

In contrast, the AA2 and AB1 stackings as shown in Fig.~\ref{fig:band_dos_mnsse}(b) and (d) favor an A-type antiferromagnetic (AAF) configuration, characterized by ferromagnetic ordering within each MnSSe monolayer and antiferromagnetic coupling between adjacent layers. In these configurations, the spin-up and spin-down band structures become nearly symmetric with respect to the Fermi level, reflecting the vanishing net magnetization of the bilayer. The stabilization of AAF ordering highlights the sensitivity of interlayer exchange interactions to the local atomic registry across the van der Waals interface, where small changes in stacking geometry can significantly modify the overlap between Mn-$d$ orbitals in neighboring layers.

The projected density of states further indicates that the electronic states near the Fermi level are predominantly derived from Mn-$d$ orbitals, with additional hybridization contributions from S-$p$ and Se-$p$ states. This orbital character suggests that the competition between ferromagnetic and antiferromagnetic interlayer coupling arises primarily from the balance between direct exchange and superexchange pathways mediated by the chalcogen atoms. Consequently, the stacking-dependent magnetic phase behavior observed in the MnSSe bilayer can be understood as a manifestation of the delicate interplay between orbital hybridization and interlayer distance inherent to van der Waals heterostructures.

Overall, these results demonstrate that bilayer MnSSe hosts multiple magnetic ground states that can be selectively stabilized through stacking control, providing a microscopic basis for tunable magnetic and spin-polarized electronic properties in two-dimensional Janus materials.

\subsection{Curie and N\'eel Temperatures of the van der Waals Bilayer MnSSe}
    \begin{figure*}[ht]
    \centering
    \includegraphics[width=13cm]{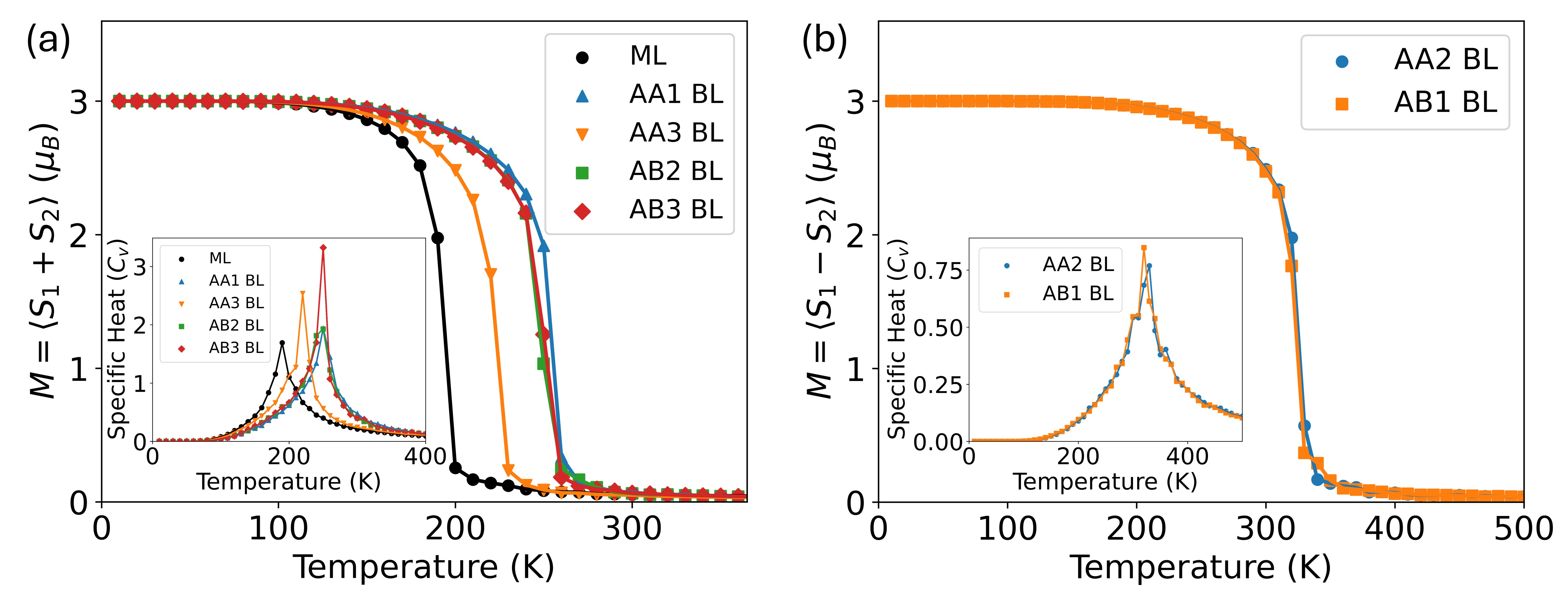}
    \caption{Temperature dependence of the magnetic order parameter for monolayer (ML) and bilayer (BL) MnSSe obtained from Monte Carlo simulations.
    (a) Curie-type ferromagnetic ordering, where the order parameter is defined as $M = \langle S_1 + S_2 \rangle$ for ferromagnetically coupled layers (AA1, AA3, AB2, and AB3 stackings). (b) Néel-type interlayer antiferromagnetic ordering for AA2 and AB1 stackings, where the staggered magnetization is defined as $M = \langle S_1 - S_2 \rangle$. The insets show the corresponding specific heat $C_v$, with the peak position indicating the magnetic transition temperature.}
    \label{fig:bilayer_Tc}
    \end{figure*}

        To investigate the magnetic phase transitions, The Monte Carlo simulations were performed using the Metropolis algorithm based on the effective spin Hamiltonian of Eqn.~\ref{MC_ML}. For the monolayer and bilayer stackings with ferromagnetic interlayer coupling ($J' > 0$; AA1, AA3, AB2, and AB3), the magnetic order parameter is defined as the total magnetization,
    \begin{equation}
    M = \left\langle S_1 + S_2 \right\rangle ,
    \end{equation}
    where $S_1$ and $S_2$ correspond to spins in the two layers. 
    As shown in Fig.~\ref{fig:bilayer_Tc}(a), the magnetization decreases continuously with increasing temperature and vanishes at the Curie temperature $T_C$. 
    The transition is accompanied by a pronounced peak in the specific heat $C_v$, which identifies the critical temperature.

    Compared with the monolayer, several bilayer stackings exhibit a significant enhancement of $T_C$. 
    This enhancement originates from the additional interlayer exchange interaction $J'$, which increases the overall magnetic stiffness of the system. Physically, the bilayer behaves as a more strongly coupled magnetic network: the intralayer exchange $J$ stabilizes ferromagnetic order within each layer, while the positive interlayer coupling $J'$ further aligns spins between layers. 
    In a mean-field picture, the critical temperature scales approximately with the total exchange field,

    \begin{equation}
    T_C \propto zJ + z'J',
    \end{equation}

    where $z$ and $z'$ denote the intra- and interlayer coordination numbers, respectively. Therefore, a positive $J'$ directly enhances the thermal stability of the ferromagnetic phase.
In contrast, AA2 and AB1 stackings exhibit antiferromagnetic interlayer coupling ($J' < 0$). 
In this case, the total magnetization vanishes even at low temperature due to compensation between the two layers. 
The appropriate order parameter is therefore the staggered magnetization,

\begin{equation}
M_s = \left\langle S_1 - S_2 \right\rangle .
\end{equation}

As shown in Fig.~\ref{fig:bilayer_Tc}(b), the staggered magnetization decreases to zero at the N\'eel temperature $T_N$, accompanied by a peak in the specific heat. 
Here, the intralayer exchange $J$ favors ferromagnetic alignment within each layer, while the negative $J'$ enforces antiparallel alignment between layers. 
The resulting ground state corresponds to a compensated bilayer antiferromagnet.

The magnitude of $J'$ plays a crucial role in determining the transition temperature. 
For large $|J'|$, the interlayer coupling significantly suppresses thermal spin fluctuations, leading to a substantial increase in the ordering temperature compared to the monolayer case.

The enhancement of magnetic transition temperatures in bilayers originates from the increased dimensionality of exchange interactions. 
While the monolayer is strictly two-dimensional, the bilayer introduces an additional magnetic coupling channel perpendicular to the plane. 
Even within a quasi-two-dimensional framework, this extra exchange pathway stabilizes long-range magnetic order and increases the critical temperature.

These results demonstrate that stacking configuration provides an effective mechanism to tune magnetic ordering temperatures in Janus MnSSe systems, enabling either enhanced ferromagnetism or compensated interlayer antiferromagnetism depending on the sign and magnitude of $J'$.

\subsection{Tunable Transition from Half-Metallic to Metallic Ferromagnetism}

    \begin{figure*}[ht]
    \centering
    \includegraphics[width=12cm]{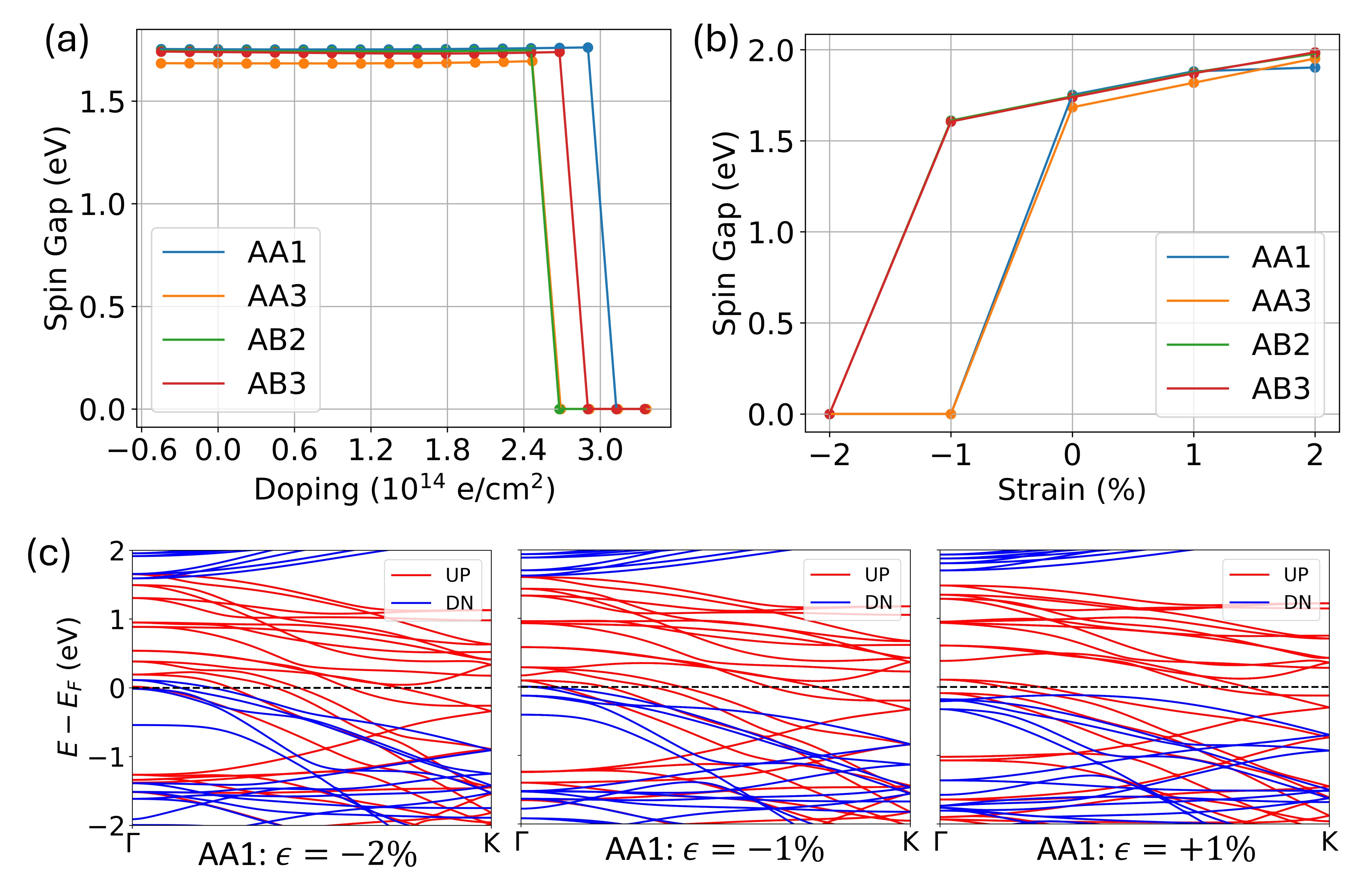}
    \caption{(a) Spin gap as a function of carrier doping for different stacking configurations (AA1, AA3, AB2, and AB3) in bilayer MnSSe. The spin gap remains nearly constant at low doping and collapses abruptly beyond a critical doping level, indicating a transition from half-metallic to metallic behavior. (b) Spin gap as a function of biaxial strain. Tensile strain enhances the spin gap, while compressive strain suppresses it, with a near-vanishing gap at large negative strain for some stackings.(c) Spin-resolved electronic band structures of the AA1 stacking under selected strain conditions ($\varepsilon = -2\%$, $-1\%$, and $+1\%$). Red and blue lines represent spin-up and spin-down channels, respectively. The evolution of the band structure illustrates the strain-driven modulation of the spin gap and the transition between metallic and half-metallic states.}
    \label{fig:doping_strain}
    \end{figure*}

    Fig.~\ref{fig:doping_strain} summarizes the evolution of the spin gap in bilayer MnSSe under external carrier doping and biaxial strain, demonstrating a highly tunable transition between half-metallic and metallic ferromagnetic states. At zero and low electron doping, all stacking configurations (AA1, AA3, AB2, and AB3) exhibit a robust spin gap of approximately $1.6$--$1.8$~eV, indicating stable half-metallic ferromagnetism. As shown in Fig.~\ref{fig:doping_strain}(a), the spin gap remains nearly constant over a wide doping range before undergoing a sudden collapse beyond a critical doping threshold. This abrupt closing of the spin gap signals a transition to a fully metallic ferromagnetic state. Notably, the critical doping level depends on the stacking configuration, suggesting that interlayer registry plays a key role in determining the electronic response to charge injection.

    A similar tunability is observed under biaxial strain, as presented in Fig.~\ref{fig:doping_strain}(b). Tensile strain enhances the spin gap, thereby stabilizing the half-metallic phase, while compressive strain reduces the gap and eventually drives it to zero. In particular, strong compressive strain leads to a near-vanishing spin gap for certain stackings, indicating a strain-induced transition to metallic ferromagnetism. This behavior originates from strain-driven modifications of orbital hybridization and exchange splitting.

    To gain microscopic insight, Fig.~\ref{fig:doping_strain}(c) shows the spin-resolved band structures of the AA1 stacking under selected strain conditions. Under compressive strain ($\varepsilon = -2\%$), bands in both spin channels cross the Fermi level, confirming a metallic state. As the strain is relaxed to $\varepsilon = -1\%$, the system approaches a critical regime where the spin gap is nearly closed. In contrast, tensile strain ($\varepsilon = +1\%$) clearly opens a gap in one spin channel while the other remains metallic, characteristic of half-metallicity. These results highlight the strong coupling between lattice deformation and spin-dependent electronic structure.

\section*{Conclusions}
In this work, we have systematically investigated the structural, magnetic, and electronic properties of bilayer Janus 1T-MnSSe using first-principles calculations combined with Monte Carlo simulations. Our results demonstrate that the magnetic ground state is strongly dependent on the stacking configuration, with the AA2 stacking stabilizing an A-type antiferromagnetic (AAF) order, indicating antiferromagnetic interlayer coupling. In contrast, other stackings favor ferromagnetic alignment, highlighting the crucial role of interlayer registry in determining magnetic interactions.

The monolayer MnSSe is confirmed to be an intrinsic ferromagnet with a Curie temperature of approximately 190 K, where magnetism is primarily governed by Mn-d orbitals. In the bilayer system, the interplay between intralayer and interlayer exchange interactions leads to enhanced magnetic stability and diverse magnetic phases, including both Curie-type ferromagnetism and Néel-type interlayer antiferromagnetism.

Importantly, several bilayer stackings exhibit robust half-metallic ferromagnetism, characterized by a finite spin gap in one spin channel and metallic behavior in the other, resulting in nearly 100\% spin polarization at the Fermi level. Furthermore, we demonstrate that the spin gap can be effectively tuned by external carrier doping and biaxial strain. A critical doping level or compressive strain drives a transition from half-metallic to fully metallic ferromagnetism, while tensile strain stabilizes the half-metallic phase.

Therefore, our findings reveal that bilayer MnSSe provides a highly tunable platform for controlling magnetic ordering and spin-dependent electronic properties through stacking engineering, charge doping, and strain. These results offer valuable insights for the design of two-dimensional spintronic devices based on Janus transition-metal dichalcogenides.

\section*{Acknowledgments}
	This research project is supported by the Second Century Fund (C2F), Chulalongkorn University (Grant No. C2F PD-2320260067). We acknowledge the supporting computing infrastructure provided by NSTDA, CU, CUAASC, NSRF via PMUB [B05F650021, B37G660013] (Thailand). (\url{URL:www.e-science.in.th}). This also work used the ARCHER2 UK National Supercomputing Service (\url{https://www.archer2.ac.uk}) as part of the UKCP collaboration.

\bibliography{references}

\end{document}